\documentclass[aps,twocolumn,prl,groupedaddress,showpacs]{revtex4-1}
\usepackage{tipa}
\usepackage{textcomp}
\usepackage{amssymb}
\usepackage{amsmath}
\usepackage[adobe-utopia]{mathdesign}
\usepackage[T1]{fontenc}

\usepackage{graphicx}

\begin{document}

\title{Self accelerating electron Airy beams}

%
\author{Noa Voloch Bloch$^{1}$}
\author{Yossi Lereah$^{1}$}
\author{Yigal Lilach$^{1}$}
\author{Avraham Gover$^{1}$}
\author{Ady Arie$^{1}$} 
\thanks{N. V. Bloch is an Eshkol scholar. This work was partly supported by the Israel Science Foundation and by the Israeli Ministry of Science. The author wishes to thank Ms. Keren Shemer and Mr. David Bloch for fruitful conversations \\}

\email[]{noavoloch@gmail.com}
\affiliation{$^{1}$ Dept. of Physical Electronics, Fleischman Faculty of Engineering, Tel Aviv University, Tel Aviv 69978, Israel\\
Author e-mail address: noavoloch@gmail.com}

\thanks{{\bf Contributions:} N.V Bloch conceived the idea.  N.V Bloch designed the experiments. N.V Bloch and Y. Lereah carried out the experiment. Y. Lilach optimized the production process and fabricated the nano-holograms.  A. Gover and A. Arie did the theoretical work and conceived ideas for applications. A.  Arie and N. V. Bloch analyzed the experimental results. A. Gover, A. Arie and Y. Lereah provided guidance.  All authors took part in writing the paper.\\
{\bf Competing Interests:} We declare that we do not have competing financial interests.}

\date{\today}
\begin{abstract}
We report the first experimental generation and observation of  Airy beams of free electrons. The electron Airy beams are generated by diffraction of electrons through a nanoscale hologram, that imprints a cubic phase modulation on the beams' transverse plane. We observed the spatial evolution dynamics of an arc-shaped, self accelerating and shape preserving electron Airy beams. We directly observed the ability of electrons to self-heal, restoring their original shape after passing an obstacle. This electromagnetic method opens up new avenues for steering electrons, like their photonic counterparts, since their wave packets can be imprinted with arbitrary shapes or trajectories. Furthermore, these beams can be easily manipulated using magnetic or electric potentials. It is also possible to efficiently self mix narrow beams having opposite signs of acceleration, hence obtaining a new type of electron interferometer.
\end{abstract}
\maketitle

An arc of light is a caustic phenomenon\cite{Berry2} abundant in nature; ranging from the rainbow in the sky\cite{Airy}, to the bright light patterns that appear at the sea floor when sun shines on the rippling waves of water\cite{Berry2}. In the framework of quantum mechanics, Berry and Balazs\cite{Berry} found a unique wave packet of a massive particle in the form of the Airy function\cite{Airy};  its counter-intuitive properties are revealed as it propagates in time or space. It preserves its shape despite dispersion or diffraction and continuously self accelerates, with no force applied, along an arc-shaped trajectory.  Nearly 30 years later, this wave packet known as \textquotesingle Airy beam\textquotesingle \; was realized by Christodoulides\cite{Siviloglou} and co-workers in the optical domain; later it was generalized to accelerating optical beams with an arbitrary spatial shape\cite{Bandres} and any curved convex trajectory\cite{Segev}. Here we report the first experimental generation and observation of the Airy beams of free electrons. The electron Airy beam is generated by diffraction of electrons through a nanoscale hologram\cite{Electron1,Electron2}, that imprints a cubic phase modulation on the beam's transverse plane. We observed the spatial evolution dynamics of an arc-shaped, self accelerating and shape preserving electron Airy beams. We directly observed the ability of electrons to self-heal\cite{selfheal}, restoring their original shape after passing an obstacle. This electromagnetic method opens up new avenues for steering electrons, like their photonic counterparts, since their wave packets can be imprinted with arbitrary shapes\cite{Bandres} or trajectories\cite{Segev}. Furthermore, these beams can be easily manipulated using magnetic or electric potentials\cite{Bliokh_Elec_1,Bliokh_Elec_2}. It is also possible to efficiently self mix narrow beams having opposite signs of acceleration, hence obtaining a new type of electron interferometer.

\begin{figure}[!!!!!!!!!!!ht]
 \includegraphics[width=81mm]{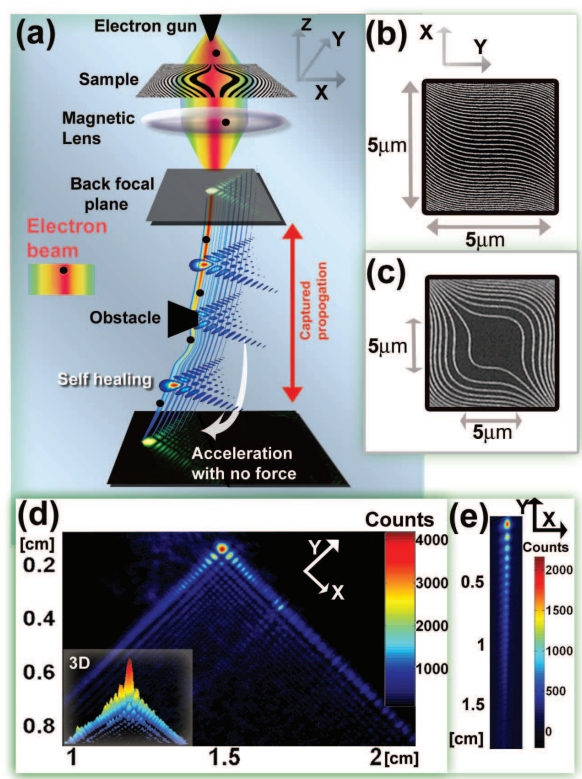}
  \caption{ {\bf  Electromagnetic generation of an electron Airy beam.} (a) An electron beam is transmitted through a nanofabricated hologram, with a cubic phase modulation. It is then focused by a magnetic lens. An electron Airy wave packet is formed at the back focal plane and recorded as it evolves. An electron Airy beam is a narrow, shape preserving, self accelerating beam. It also self-heals after encountering obstacles. (b, c) TEM micrographs of the nanoscale holograms (b) 2D Airy on a spatial carrier frequency. (c) 2D Airy without a carrier. (d, e) Experimental wave packet micrographs of 2D and 1D electron Airy beams.
}\label{fig:Airyexp}
\end{figure}
When a particle's wave packet (representing its probability density), is given the initial shape of an Airy function, it does not spread and continuously self accelerate as it evolves in time. This was theoretically predicted by Berry and Balazs in 1979\cite{Berry} as they showed that the Airy wave packet is a solution to the Schr\"{o}dinger equation for a free particle.     
The evolution of a slowly varying amplitude light beam in space is analogous to the evolution of the wave packet of a massive particle in time; much as the paraxial Helmholtz equation resembles the Schr\"{o}dinger equation. This analogy led Christodoulides and co-workers\cite{Siviloglou} to discover and experimentally realize the \textquotesingle accelerating Airy optical beams\textquotesingle, the optical analog of the quantum Airy wave packet of a massive particle\cite{Berry}.  The accelerating Airy wave packet does not contradict the Ehrenfest theorem as its center of mass propagates in straight lines, but the highest probability amplitude of the particle's location, i.e., the strongest lobe of the Airy function preserves its shape and stays localized around a parabolic trajectory in space, similar to that of a freely propagating projectile experiencing a transverse accelerating force.  This intriguing propagation dynamics is a caustic wave phenomena which can be understood by ray analysis: multiple rays  emerge sideways from the edge lobes of the initial wave packet area and coalesce along a curved boundary\cite{Yan}. 
Various applications followed the discovery of optical Airy beams including micro-particle manipulation\cite{Paticle_manipulation}, generation of plasma channels in air and water\cite{plasma}, surface Airy plasmons\cite{Salandrino, Minovich, zhang} and applications in lasers\cite{Gil, Longhi} and in nonlinear optics\cite{Ellenbogen}. However, all these applications relied on the wave-packet of zero mass photons. We report here the first generation and observation of Airy wave packet\cite{Berry} of free electrons, or \textquotesingle electron Airy beams\textquotesingle\; (generated without using photons). We demonstrate a new electromagnetic technique to generate electron Airy beams analogous to the optical method\cite{Siviloglou}, as we exploit the recent advances in nanoscale hologram fabrication techniques\cite{Electron1,Electron2}.
We supply a robust method to manipulate the electrons by trajectory, as their high peak intensities (indicating the probability density for locating the electron) self accelerate in a parabolic path. Furthermore, this concept can be generalized because recent studies have shown that optical Airy beams can be designed with  arbitrary spatial shapes\cite{Bandres} or propagate along arbitrary trajectories\cite{Segev}.

\begin{figure}
\includegraphics[width=88mm]{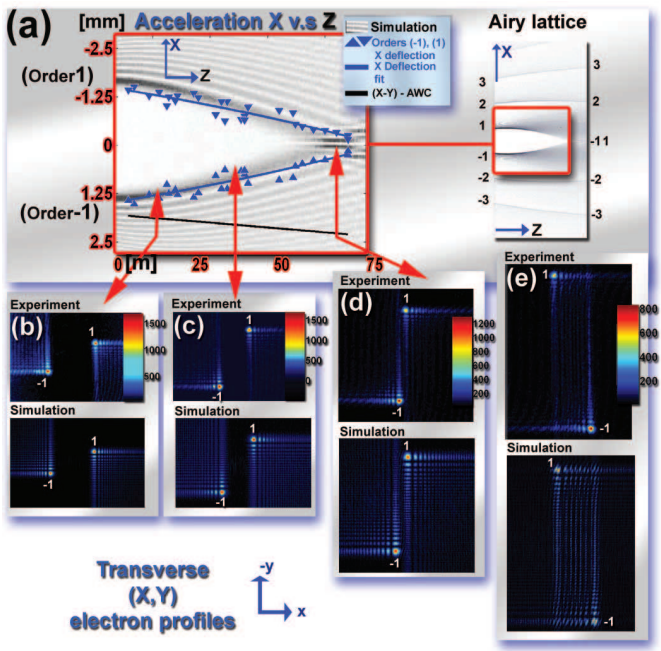}
  \caption{{\bf Acceleration of Airy beams}. (a) The measured trajectories of Airy orders (1,-1) (blue curves) fits the simulation (gray background). Airy orders accelerate anomalously inward (to the center of the axis). However the carrier (black curve) deflects in the usual manner outward with a constant transverse velocity. (b-e) Micrographs revealing the evolution and acceleration of electron Airy beams v.s numerical simulation results (d) Airy orders can collide (in x axis), (e) or pass each other. Notice that the peak intensities of the different orders rotate clockwise (e.g. for the -1 order), or anti clockwise (for the +1 order).}
\label{fig:results}
\end{figure}

The spatial evolution of the envelope $\Psi$ of an electron's wave function, can be expressed by the paraxial Helmholtz equation (see Supplementary Information Sec. 1):
\begin{equation}{
\left(\nabla_\bot^2+2ik_B\frac{\partial}{\partial z}\right)\Psi=0}\;\;,
\end{equation}
where ${\nabla_\bot^2=\partial/x^2+\partial/y^2}$ is the transverse derivative and ${k_B=p/\hbar=2\pi/\lambda_B}$ is the de-Broglie wave number of the electron. This equation has the same form as that of the Schr\"{o}dinger equation. However, rather than measuring the evolution of the envelope of the electron wave function in time, we measure it in space. This is another manifestation of the analogy, widely used in optics, between beam diffraction in space and a pulse dispersion in time. 
When the initial probabilistic wave function of the electron is an Airy function (Ai), ${\Psi(x,y,z=0)=\mathrm{Ai}(x/x_0)\cdot \mathrm{Ai}(y/y_0)}$, where ${x_{0}, y_{0}}$ are characteristic length scales; the general solution to the wave packet is given by\cite{Berry, Paticle_manipulation, selfheal}:
\begin{widetext}
\begin{equation}
\Psi(x,y,z)= \mathrm{Ai}\left(\frac{x}{x_0}-\frac{z^2}{4k_B^2x_{0}^4}\right)\cdot \mathrm{Ai}\left(\frac{y}{y_0}-\frac{z^2}{4k_B^2y_{0}^4}\right)
\cdot \exp\left(i\frac{xz}{2k_Bx_0^3}-i\frac{z^3}{12k_B^3x_0^6}\right)\cdot \exp\left(i\frac{yz}{2k_By_0^3}-i\frac{z^3}{12k_B^3y_0^6}\right)\;\;.
\label{eq:funcAiry}
\end{equation}
\end{widetext}
It is then clear from Eq. \ref{eq:funcAiry} that $|\Psi|^2$ preserves its shape and accelerates in the transverse coordinates. The acceleration is described by\cite{christ} ${x(z)=z^2/4k_B^2x_0^3}$, ${y(z)=z^2/4k_B^2y_0^3}$. 
The aforementioned  ideal Airy beam carries an infinite amount of energy, wheres the feasible Airy beam is truncated, having a finite energy. The finite Airy beam is obtained by multiplying the Airy function with an exponential or Gaussian window\cite{Siviloglou}. Nevertheless, over a finite distance, the finite Airy beam exhibits all the special characteristics of the infinite Airy beam such as low diffraction, free acceleration and self-healing.

In optics, finite Airy beams were experimentally obtained by passing a Gaussian beam through a phase mask imprinting a cubic phase modulation  in the transverse direction\cite{Siviloglou}, followed by an optical Fourier transform
(notice that the Fourier transform of a function having a cubic phase modulation results in an Airy function). In our experiment we utilize a field emission gun transmission electron microscope (FEG-TEM), operating at $\mathrm{200 \; kV}$. The de-Broglie wavelength (including relativistic correction) in this case is approximately $\mathrm{2.5\;pm}$. Therefore to view the inner structure of Airy beams as well as the spatial separation of different orders,
we used nanoscale holograms (see Supplementary information Sec. 2). The emitted electrons pass through the holograms which adds a transverse cubic phase:\\ ${\exp(i\phi(x,y))=\exp(ia_{x}x^3)\cdot\exp(ia_{y}y^3)}$, to the wave function. 
Our hologram design method was to construct a binary diffraction grating with the following shape\cite{binaryAiry, Lee}:
\begin{equation}{
h(x,y)=\frac{1}{2}h_0(\mathrm{sgn}\{\cos[2\pi x/\Lambda+a_xx^3+ a_yy^3]+D_{cycle}\}+1).} \label{eq:holo}
\end{equation}
In this way, a cubic phase is imposed on a carrier frequency. The carrier period is $\Lambda$, $h_0$ is the ridge height of the binary phase mask and ${D_{cycle}}$ is an arbitrary duty cycle factor. 
The micrographs of these nanoscale holograms are shown in Fig. \ref{fig:Airyexp}b, c. The most important holograms for our experiment are: A 2D Airy lattice structure with carrier (AWC): ${\Lambda_X=400\;{\mathrm n\mathrm m}}$, $a_{x,y}={2\cdot 10^{10}\cdot 2\pi/(400\cdot 10^{-9})}$ ${[1/\mathrm m^3]}$  shown in Fig. \ref{fig:Airyexp}b. A simple periodic Bragg lattice (BR) structure with a period ${\Lambda_X=100\;\mathrm n\mathrm  m}$.
Fourier transform of the modulated wave function is done by using a set of magnetic lenses, so that an electron Airy beam is obtained at the back focal plane of the FEG-TEM. This method is analogues to the method used to generate optical Airy beams\cite{Siviloglou}, and the only difference is that we manipulate electrons, rather than photons. The experimental Airy profiles of 2D and 1D Airy beams are shown in Fig.  \ref{fig:Airyexp}d, e. Since the measurement plane is located at a fixed position in our FEG-TEM, we varied the focal lengths of the magnetic lenses in order to observe the formation and evolution of the electron Airy wave packet in space as illustrated in Fig. \ref{fig:Airyexp}a. The experiment details are given in Supplementary information Sec. 3.

In the first experiment we measured the acceleration of electron Airy beams (results displayed in Fig. \ref{fig:results}). We recorded profiles using a relatively large area ${(\mathrm 100 \;\mu \mathrm m \; X \; 100\; \mu \mathrm m)}$. The diffraction pattern from an AWC includes several diffraction orders. The zero diffraction order is not an Airy beam,  therefore it rapidly expands and becomes invisible at the measurement plane.  However, all the higher order modes are accelerating and shape preserving (hence spatially localized) Airy beams. The acceleration of these beams can be shown in the x-z plane or the transverse x-y plane as presented in Fig. \ref{fig:results}a, (b-e) respectively. We simulated the evolution of the electrons, by using beam propagation numerical simulation\cite{Ellenbogen}. The numerical results are in good agreement with the experimental results. Detailed explanation of acceleration measurement  is given in Supplementary Information Sec. 4. In particular, it can be seen that the Airy beams of order +1 and -1 accelerate toward each other, eventually colliding, as shown in  Fig. \ref{fig:results}d, e. This represents a new way of interfering electron beams: The electron beam is separated into two diffraction orders, but these beams re-merge owing to their opposite acceleration directions. The measured deflection is 1.7 mm over an effective distance of 70 m. Notice that  the acceleration can also be viewed in a single plane, as the observed Airy orders reach different transverse heights (see Supplementary Information Sec. 5). 

\begin{figure}[!!!!h]
\centering
 \includegraphics[width=88mm]{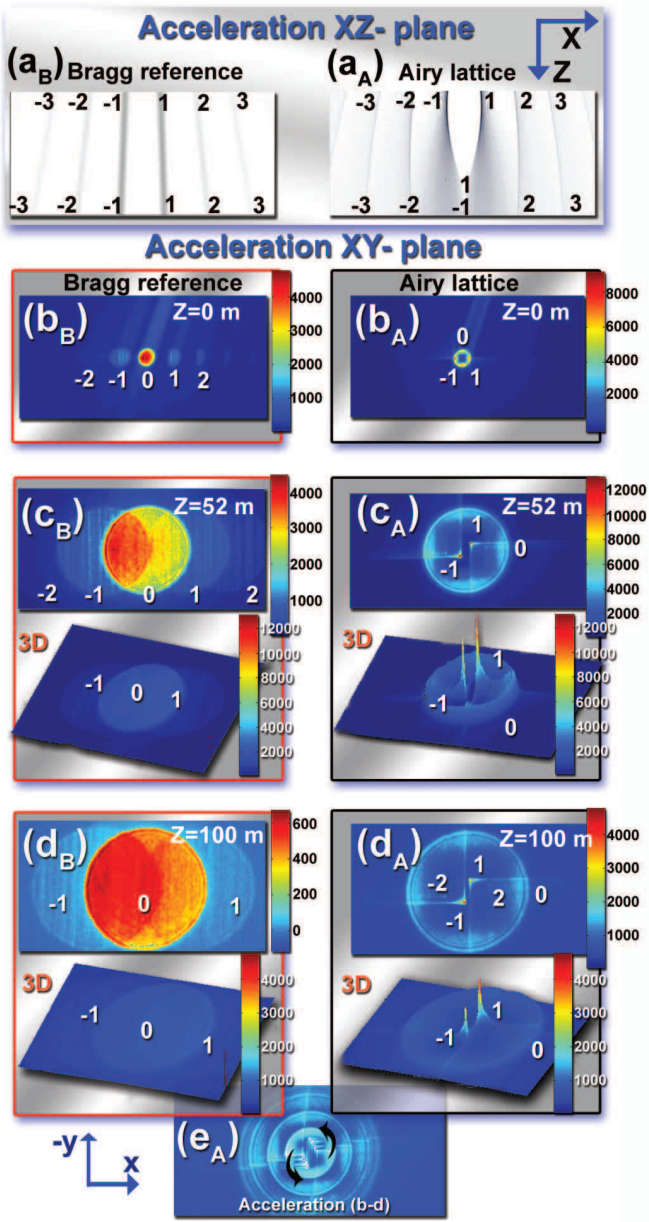}
  \caption{{\bf Comparison between electrons diffracting from an Airy lattice to the electrons diffracting from a reference periodic Bragg grating.} (a) The diffraction from a Bragg lattice is normally outward at angles\cite{Electron2}: ${\alpha_m=m\lambda_B/\Lambda}$.  However, the diffraction from an Airy lattice, is anomalous because the lattice peaks accelerate inward (b-d) Experimental profile micrographs of different propagation planes.  (notice that in both the  Airy lattice and the Bragg grating the zero order looks the same). The Airy orders are very localized and maintain very high intensities compared with the Bragg orders. (e) An overlay of different profile micrographs.}\label{fig:Airybroaden}
\end{figure}

\begin{figure}[!!!!h]
\begin{center}
 \includegraphics[width=88mm]{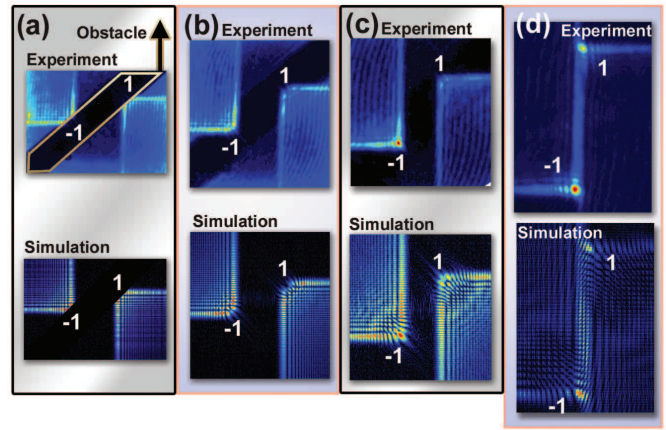}
  \caption{{\bf Self healing properties of electron Airy wave packets. }Experimental profile photograph versus numerical simulation. The diffracted orders 1 and -1 are blocked differently so their self-healing progress is different. The two orders are blocked with a glass wire.}\label{fig:Selfhealing}
\end{center}
\end{figure}

In the second experiment, we compared between electrons diffracting from an Airy lattice to the electrons diffracting from a reference periodic Bragg grating. We simulated the evolution of electrons (Fig. \ref{fig:Airybroaden}a) diffracted from a periodic Bragg lattice and an Airy lattice with the same carrier period. (Notice that in this case only, we used different input beam area for Airy and Bragg lattices, in order to visualize the trajectories). As seen, the diffraction from a Bragg lattice is normally outward, however, the diffraction from an Airy lattice, is anomalously accelerating inward. We recorded profiles (Fig. \ref{fig:Airybroaden}b-d) using a relatively small area ${(\mathrm 10\;\mu \mathrm m \; X \; 10\; \mu \mathrm m)}$, letting the zero order to slowly spread out thereby enabling visualization of the ratio between the evolution of spreading Bragg peaks compared to the shape-preserving Airy lattice peaks.
The zero order of Airy lattice is the only diffraction order, which is not imprinted with a cubic phase, thus, it spreads in the normal manner. The zero order appears as a large circular spot at the middle of the frame.
We can also see diffraction orders 1, -1, 2, -2. Notice that the intensity of Bragg peaks with opposite signs is not symmetric as expected. This is probably because the sample was slightly tilted and not perfectly perpendicular to the electron beam.  As can be seen, at an effective distance of 52 m and 100 m the diffraction patterns from BR (as well as from the zero order of AWC) spread and become huge, while the non-zero Airy peaks stay confined along a very large effective distance (even after an effective distance of 100 m) to a very small scale area. This difference is emphasized when calculating the full width half maxima (FWHM) of the diffracted orders. FWHM of Bragg orders (1,-1) are: $\mathrm{(b),\; (c),\; (d)= 1125,\; 5785,\; 8125\;\mu m}$, however,  FWHM of Airy orders $\mathrm{(1,-1)}$ are:  $\mathrm{(b),\; (c),\; (d)=102,\; 104, \;110 \;\mu m}$. We can compare the FWHM of the diffracted orders: The ratio between the FWHM, i.e.,  $\mathrm{FWHM (Bragg)/FWHM(Airy)}$ is about: 11, 58, 81 respectively.
Notice that the profiles resemble the profiles given in Fig. \ref{fig:results} but the Airy function in this case, is more truncated. 

We also measured the self healing properties of electron Airy beams, as presented in Fig. \ref{fig:Selfhealing}.  For this purpose, we used a wire placed in the diffraction plane. The wire was conventionally used as a bi-prism in electron holography\cite{Lichte}, but in our case it was simply used to block parts of the beam. This experiment was conducted in the following way: increasing the current of the  objective lens elevated the Airy beam above the wire. Then we adjusted the wire to simultaneously block the two Airy orders of -1 and 1. We blocked the different orders in a slightly different manner. We then gradually increased the current of the diffraction lens and observed their self-healing progress\cite{selfheal}. The wave packets reconstructed their shapes after passing the blocking wire. As shown, the healing progress of the orders was different. Order 1 self-healed faster than order -1.  We also simulated the self healing process and the numerical results are in a good agreement with the experimental results.

We demonstrated experimentally, for the first time, non spreading, freely accelerating electron wave packets in the shape of an Airy function. This gives rise to a novel field of manipulating particles' trajectories by engineering their probability density wave functions. We experimentally demonstrated trajectories of electrons, accelerating in free space without induction of force. We observed shape preserving, highly localized electron Airy beams which didn't spread even after an effective length of 100 m. Such non-spreading Airy electron wave packets may possibly be useful for improving the resolution properties of TEM imaging, since they have an extremely large depth of focus and a resolution depth feature. Furthermore, we showed, for the first time, that these self accelerating electron beams can self heal and reform their original shape after passing an obstacle. We demonstrated an interesting feature of these beams, in which different diffraction orders with opposite signs of acceleration can be merged and possibly be used as a new type of electron wave interferometer. It may also be interesting to study with these beams electron spin interaction effects in the relativistic regime of the electron wave-packets, similarly to recent studies of electron vortices\cite{Bliokh_Elec_2}. New possibilities may open up for manipulating and shaping the trajectories and the self healing properties of Airy beams, as electrons can be influenced by magnetic or electric potentials\cite{Bliokh_Elec_1}.

\section{Supplementary Information}
\subsection{1. Quasi-relativistic  Schr\"{o}dinger equation} 
Ignoring spin effects, we may use the Klein-Gordon equation instead of Dirac equation in order to find the wave function of an relativistic electron. For free space propagation (without any external potential):
\begin{equation}{
-\hbar^2\frac{\partial^2}{\partial t^2}\Phi=(mc^2)^2\Phi-c^2\hbar^2{\nabla^2}\Phi}.\label{eq.KG}
\end{equation}
Its electron wavefunction plane wave solution is:
\begin{equation}{
\Phi(r,t)=\Phi_0\exp(i \underline{p}\cdot\underline{r}/\hbar-i Et/\hbar)\;\;,}
\end{equation}
where E, p satisfy the dispersion relation ${E=\sqrt{c^2p^2-(mc^2)^2}}$.
Classically, ${E=\gamma mc^2}$, ${P=\gamma \beta m c}$, ${\gamma=1/\sqrt{1-\beta^2}}$ and ${\beta=v/c}$. \\
For small angle diffraction ${p_{\bot}\ll p}$ (paraxial approximation) we can look for a wave solution of the form:
\begin{equation}
{{\Phi(r,t)=\Psi(r_{\bot},z)\cdot \exp(i Pz/\hbar-iEt/\hbar)}\;\;,}
\end{equation}
Assuming slowly varying envelope and neglecting the second derivative of z, from Eq. \ref{eq.KG} we obtain the paraxial Helmholz equation:
\begin{equation}{
\left(\nabla_\bot^2+2ik_B\frac{\partial}{\partial z}\right)\Psi=0}\;\;,
\end{equation}
where ${\nabla_\bot^2=\partial/x^2+\partial/y^2}$ is the transverse derivative and ${k_B=p/\hbar=\gamma mv/\hbar=2\pi/\lambda_B}$ is the de-Broglie wave number of the electron.
This equation has the same form as the paraxial Helmholtz equation, which describes the propagation of light beams in space (used to find the solution for Airy optical beams\cite{Siviloglou}), except that it includes the de-Broigle wave number.

Setting ${z=\beta ct}$  results in:
\begin{equation}{
\left(\frac{\hbar^2}{2\gamma m_e}{\nabla_\bot^2}+i\hbar\frac{\partial}{\partial t}\right)\; \Psi=0\;\;,}
\end{equation}
which is the same as the Schroedinger equation that was solved (in 1D) by Berry and Balzas\cite{Berry}, with a relativistic correction to the electron rest mass.\label{S3}
\subsection{2. Nanoscale hologram preparation}
Our hologram design method was to construct a binary diffraction grating with the following shape\cite{binaryAiry, Lee}:
\begin{equation}{
h(x,y)=\frac{1}{2}h_0(\mathrm{sgn}\{\cos[2\pi x/\Lambda+a_xx^3+ a_yy^3]+D_{cycle}\}+1).} \label{eq:holo}
\end{equation}
In this way, a cubic phase is imposed on a carrier frequency. The carrier period is $\Lambda$, $h_0$ is the ridge height of the binary phase mask. ${D_{cycle}}$ is an arbitrary duty cycle factor. Setting the duty cycle factor $0<D_{cycle}<1$ to high values (0.8 , 0.9...), gives a hologram with very narrow stripes. This method was chosen because it matches well with the highest resolution writing mode of the focused ion beam (FIB).
The nanoscale holograms were prepared by sputter depositing 10 nm Au layer onto a 50 nm silicon membrane chip, then milling the desired pattern with a Raith IonLine focused ion beam machine. The machine mills with a 35 keV Ga ions beam, and we used 2.6 pA ion beam current with step size of 2nm and dwell time of 1.54 msec.  This gives an effective dose of 20000 pC/cm  for single-pixel-line milling.  These conditions milled through the entire gold layer and in addition about 20 nm of the silicon nitride membrane. The milling was done over areas of $30\;\mu \mathrm{m}\;X\;30\;\mu \mathrm{m}$ as shown (only in the central part) in Fig. 1b, c.\label{prep}. This method enabled us to achieve high resolution writing, without breaking the membrane chip. This kind of resolution was necessary for our experiment.

\begin{enumerate}
\item A 2D Airy lattice structure with carrier (AWC): ${\Lambda=400\;\mathrm{nm}}$ and $a_{x,y}={2\cdot 10^{10}\cdot 2\pi/(400\cdot 10^{-9})}$ ${[1/\mathrm{m^3}]}$  shown in Fig. 1b.
\item A 2D structure with no carrier (ANC):
$a_{x,y}={2\cdot 10^{10}\cdot 2\pi/(400\cdot 10^{-9})}$  with the same acceleration as AWC ${[1/\mathrm{m^3}]}$ as shown in Fig. 1c.
\item A 1D structure with carrier:
${\Lambda_X=400\;\mathrm{nm}}$, and acceleration only in y: $a_{y}={2\cdot 10^{10}\cdot 2\pi/(400\cdot 10^{-9})}$ ${[1/\mathrm{m}^3]}$ .
\item A simple periodic Bragg lattice (BR) structure with a period ${\Lambda_X=100\;\mathrm{nm}}$. It served as a reference scale for the measurement.
\end{enumerate}\label{S2}
\begin{figure}[h!!!!]
\centering
 \includegraphics[width=88mm]{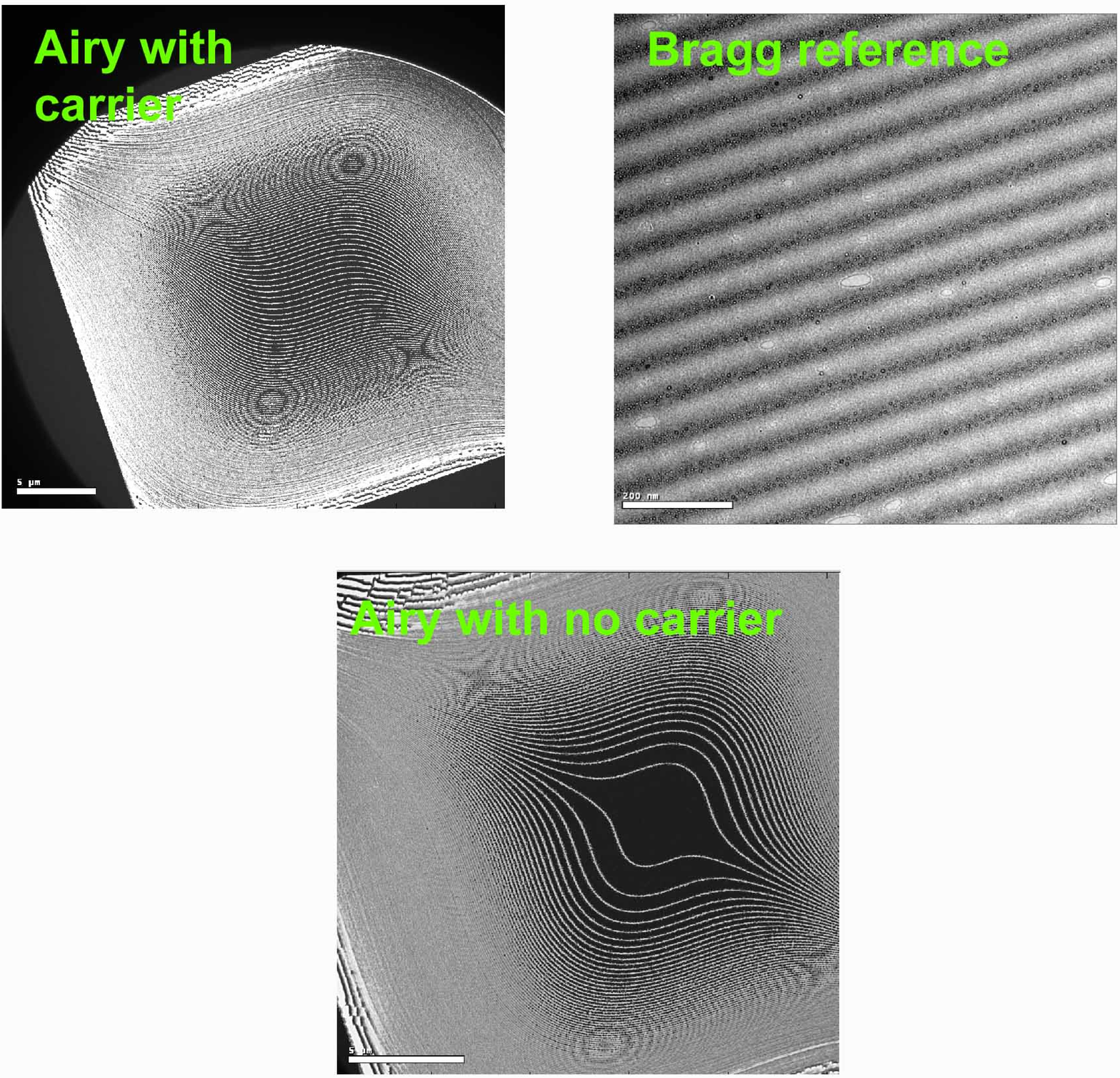}
  \caption{{(color online)}. Micrograph of AWC, ANC BR. The aforementioned method of writing  enabled us to achieve high resolution, without breaking the membrane chip. This kind of resolution was necessary for our experiment.}
\end{figure}
\label{S4}

\subsection{3. Experiment details}
We used a Tecnai F20@ FEG-TEM and studied the generation and evolution of electron Airy beams by varying the focal lengths of the TEM magnetic lenses\cite{Electron2}. We imaged the electron wave packet as it evolved in different planes between the back focal plane and the image plane. The profiles were taken in FEG-TEM low angle diffraction mode (LAD), for which the objective lens is with low current (around 7 \% of the maximal current). This enabled high magnification of the back focal plane. Setting a high current in the magnetic condenser lens and using an area adjuster, gave a relatively low convergence electron beam. Then, by a  \textquotesingle Free Lens Control\textquotesingle \; software increasing the current in the diffraction magnetic lens only (which was the only lens we changed during the measurements), we imaged different planes following the back focal plane as shown in Fig. 1a. A Gatan camera model 694 captured the wave-packet evolution and the spatial propagation dynamics of the  electron Airy beam.
\label{S5}

\subsection{4. Acceleration measurement}
\begin{figure}[h!!!!]
\centering
 \includegraphics[width=88mm]{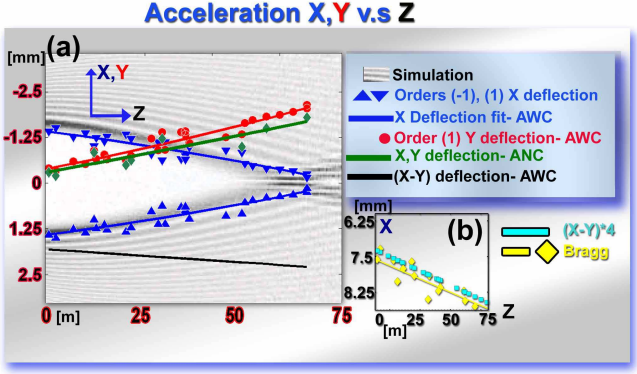}
  \caption{{(color online)}. For measuring the acceleration we used three structures : AWC, periodic Bragg reference (BR)  and additional structure only having an Airy modulation with no carrier [Airy with no carrier (ANC)]. We have also recorded the electron evolution from ANC and BR for same z (at the identical experimental parameters) and analyzed the results.structure\label{fig:Accelerationmes}}
\end{figure}
One of the challenges in this setup is to determine the effective distance that the beam travels from the focal plane. We change this distance by decreasing the focal length of the magnetic lens each time. In
order to calibrate the propagation (z) axis we made the following analysis:  We measured the Bragg diffraction patterns from the periodic reference grating for different setting of the magnetic lens. The location of the first diffraction order is given by\cite{Electron2}: 

\begin{equation}{
Peak_{x(BR)}(z)=x_{s(BR)}+\frac{\lambda_B}{\Lambda_{x(BR)}}z.}
\end{equation}

This diffraction is measured using the camera, and since each pixel size is 25 microns, we can obtain the distance z in meters. As for the diffraction of the Airy wave through the mask with carrier (AWC), it has both acceleration and linear velocity components in x direction, i.e. the peak location of the strongest lobe is given by \cite{christ}:

\begin{equation}{
Peak_{x(Ai)}(z)=x_{s(Ai)}+\frac{\lambda_B}{\Lambda_{x(Ai)}}z+\frac{1}{4k_B^2x_0^3}z^2}\;\;,
\end{equation}
whereas in the Y direction we have only acceleration, hence the peak location is:
\begin{equation}{
Peak_{y(Ai)}(z)=\frac{1}{4k_B^2x_0^3}z^2\;\;.}
\end{equation}
Therefore, the difference between then is:
\begin{equation}{
Peak_{x(Ai)}(z)-Peak_{y(Ai)}(z)=x_{s(Ai)}+\frac{\lambda_B}{\Lambda_{x(Ai)}}z\;\;,}
\end{equation}
which is also linearly increasing with a constant transverse velocity in z. Since the period here is ${\Lambda_{x(Ai)}=400\;\mathrm{nm}}$, vs. only ${\Lambda_{x(BR)}=100\; \mathrm{nm}}$ for the Bragg reference mask, the slope and the starting point ${x_s}$ in the case of the Airy beam should be 4 times lower. We can therefore verify the calibration by comparing the two measurements, as shown in Fig. \ref{fig:Accelerationmes}b.

We can now compare the measured peak location in the X axis, when diffracted from the AWC mask with the theoretical prediction. The +1 and -1 diffraction orders are clearly seen, and they are accelerating toward the center. There is an excellent agreement with the theoretical prediction.
In addition, we have the results from this mask in the Y direction of orders +1 and -1. Since there isn't a carrier modulation in the mask's Y direction, the beam starts in this case from the axis and accelerates away from it, as shown in the red curve of Fig. \ref{fig:Accelerationmes}a. We have also measured the diffraction patterns from the ANC mask. Here there is no carrier modulation, neither in x nor in y, hence the beam starts from the axis and accelerates away from it (green line in Fig. \ref{fig:Accelerationmes})a. As expected, it nearly coincides with the results of the Y axis from the AWC mask. The measured deflection is 1.7 mm (~70 pixels) over an effective distance of 70 m. 
\label{S1}

\subsection{5. The Airy lattice}

\begin{figure}[h!!!!]
\centering
 \includegraphics[width=88mm]{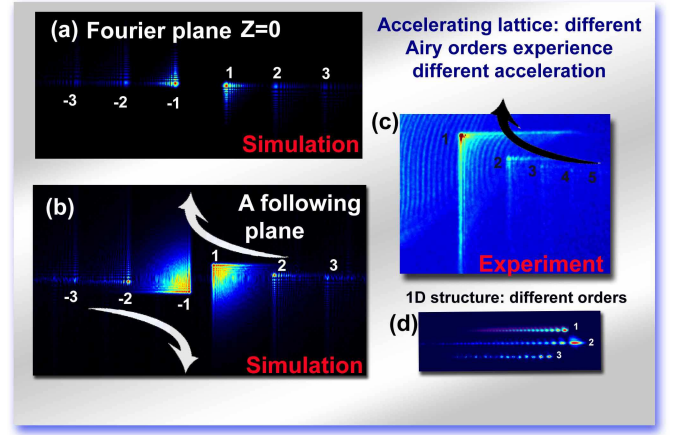}
  \caption{{(color online)}. Localized accelerating lattice. The acceleration can be also viewed in one profile picture taken after the back focal plane. Different orders have different acceleration coefficients. (a) Simulation of Fourier plane of an Airy lattice. (b) Simulation of a following plane. As can be seen each order elevate to a different height (in y). (c) Experimental profile showing this effect. (d) Airy orders have also different spatial distribution as shown with the different orders of the 1D structure\label{fig:AiryLattice}}
\end{figure}
Currently, the acceleration measurement is done by taking several profile pictures in different plains following the Fourier plane. The purpose of this section is to explain how the acceleration in the Airy lattice can 
be indicated by single plane (after Fourier plane).
  
The Airy lattice is a periodic lattice imposed with a cubic phase. 
A general formula to impose a lattice with an arbitrary phase is (in binary structures) \cite{Lee}:
\begin{equation}{
h(x,y)=\frac{1}{2}h_0\cdot sgn(\cos[2\pi x/\Lambda+\phi(x,y)]).} \label{eq:holo}
\end{equation}
Any arbitrary phase can be imposed on this lattice. As an example the fork shaped structure\cite{Electron1,Electron2} is generated by: $\phi(x,y)=l_c\tan^{-1}(y/x)$, where $l_c$ is the topological charge. 
The Airy lattice is imposed with following phase:
$\phi(x,y)=a_xx^3+a_yy^3$.
When electrons (or light) diffracts from the aforementioned binary structure,
it decomposes to different orders; the complex amplitude of the m-th order diffracted beam is $\propto \exp(i\cdot m \phi(x,y))$. 
In the case of fork shape structure, series of vortices is obtained. Each vortex  has a different orbital angular momentum, which  equals to $m\cdot l_c$. The radius of each vortex increases as a function of orbital angular momentum, thus, the diffraction from a fork shaped structure generates a series of vortices with different radii\cite{Electron1,Electron2}. 
In the case of Airy lattice, each diffracted order is imposed with a different 
cubic phase:  $\propto \exp(i\cdot m (a_xx^3+a_yy^3))$, thus, each order experience different acceleration. This can be viewed in a single plane as shown in Fig.  \ref{fig:AiryLattice} it was also observed experimentally  Fig. \ref{fig:AiryLattice}c 

\label{S4}
This is a novel type of lattice. Although it is decomposed to different orders which propagate in different directions, each single order stays localized (unlike Bragg lattice) and anomalously accelerate.

\end{document}